\documentclass[a4paper,11pt]{article}
\usepackage{pos}
\usepackage{bbold}
\usepackage{amsmath}
\usepackage{booktabs}
\usepackage{lineno}
\usepackage{lipsum}
\usepackage{color}
\usepackage{hyperref}
\usepackage[nameinlink]{cleveref}
\usepackage{graphicx}
\usepackage{dcolumn}
\usepackage{slashed}

\title{Low energy effective theories on the lattice with coloured noise}

\author*[a]{Felipe Attanasio}
\author[a]{Jan Philipp Klinger}
\author[a,b]{Jan M.~Pawlowski}

\affiliation[a]{Institute for Theoretical Physics, Universit\"at Heidelberg,\\
  Philosophenweg 16, D-69120, Germany}

\affiliation[b]{ExtreMe Matter Institute EMMI, GSI,\\
Planckstra{\ss}e 1, D-64291 Darmstadt, Germany}

\emailAdd{pyfelipe@thphys.uni-heidelberg.de}
\emailAdd{klinger@thphys.uni-heidelberg.de}
\emailAdd{j.pawlowski@thphys.uni-heidelberg.de}

\abstract{
Low energy effective theories give access to regimes of the QCD phase diagram that to date are hard to simulate directly with lattice QCD or with functional approaches. For lattice QCD this includes the small temperature and/or large density regime. In both regimes the lower UV cutoff in low energy effective theories may soften computational problems. Moreover, lattice results for low energy effective theories serve as benchmark results for  functional approaches for these effective theories. Here we present lattice results for the scalar O($4$) and quark-meson models. We simulate the theory via Stochastic Quantisation and report on the effects of employing coloured noise, a method that allows control over the momentum scale of the simulation.}

\FullConference{%
  The 39th International Symposium on Lattice Field Theory (Lattice2022),\\
  8-13 August, 2022 \\
  Bonn, Germany 
}

\begin{document}
\maketitle

\section{Introduction}

Low energy effective theories (LEFTs) for strongly interacting matter allow for an understanding of the mechanisms behind physical processes in QCD. They can be embedded into QCD within integrating out part of the fundamental degrees of freedom, while keeping the dynamical low energy degrees of freedom. In most cases this is done by fully integrating out the gluons and the high energy quark momentum modes. The remaining effective theories contain dynamical quarks and the dynamical light hadrons. 

In short, LEFTs facilitate a qualitative or semi-quantitative analysis of challenging problems, that to date cannot be resolved within first principles approaches to QCD: For instance, lattice LEFTs allow for simulations at smaller temperatures without large temporal extents due to their small physical UV-cutoff. Potentially, this gives an access to regions of the QCD phase diagram that are inaccessible to full QCD simulations. Moreover, LEFT actions, still coupled to glue dynamics, emerge dynamically in the Wilsonian functional approach at low momentum scales, see the functional renormalisation group (fRG) reviews  \cite{Dupuis:2020fhh,Fu:2022gou}. Hence, lattice results for these LEFTs provide benchmark tests for this part of first principle functional computations in QCD, as well as functional computations directly in LEFTs.   

\section{Quark-meson model}

The Quark-Meson (QM) model is a commonly used low energy effective theory of QCD. Its dynamical degrees of freedom are the quarks and the light scalar and pseudo-scalar mesons, and its physical ultraviolet cutoff is of the order of 1\,GeV. Its classical action is given by 
\begin{align}\nonumber 
S_{\textrm{QM}}[q,\bar q,\phi] = &\,\int d^d x\,\Biggl\{ \bar q \left[ \slashed{\partial}  + \frac{h_\phi}{2}\left( \sigma + i \gamma_5 \vec{\tau}\cdot\vec{\pi}\right) \right] q \\[2ex] 
& \hspace{1.5cm}+ \frac12\left[ (\partial_\mu\sigma)^2+ (\partial_\mu\vec\pi)^2 \right]+ \frac12 m_\phi^2(\sigma^2+\vec \pi^2) + \frac{\lambda_\phi}{8} (\sigma^2+\vec \pi^2)^2  \Biggr\}\,,
\label{eq:QMAction} 
\end{align}
with the quark field $q$ comprising the up and down quarks, $q^T=(u,d)$, and the scalar fields are associated with the $\sigma$ and $\vec{\pi}$ mesons. In \labelcref{eq:QMAction} we have also used $\vec \tau^T =(\sigma_1,\sigma_2,\sigma_3)$ with the Pauli matrices $\sigma_i$. The coupling $\lambda_\phi$ introduces meson scattering events, and the Yukawa coupling $h_\phi$ describes the scattering of a quark--anti-quark pair into a meson. The action in \labelcref{eq:QMAction} is an $O(4)$-invariant low energy effective theory with $N_f = 2$ quark flavours, $N_c = 3$ colours and four mesonic degrees of freedom. 

The QM model takes into account the low energy effects of the scalar-pseudoscalar channel of the four-quark scattering in QCD. This channel has been shown to be dominant in the vacuum, \cite{Mitter:2014wpa,Cyrol:2017ewj}, and at sufficiently small chemical potential, \cite{Braun:2019aow}. In short, it captures well the chiral low energy dynamics of QCD. Its emergence from full QCD at finite density has been quantitatively studied in \cite{Fu:2019hdw}, and its validity range, that is its ultraviolet cutoff, has been studied in \cite{Rennecke:2015lur,Alkofer:2018guy}. 

The QM model and its variants or extensions are still commonly used within many applications to the phase structure of QCD, including applications at large densities, isospin chemical potential or strong magnetic fields. For a compilation of various applications as well as respective results see the recent review \cite{Dupuis:2020fhh}.  

We now proceed with the lattice setup of the quark-meson model with the action \labelcref{eq:QMAction} in $d$ Euclidean spacetime dimensions. We define 
\begin{align}
    S &= \sum_{x,y} \bar{q}_x \overbrace{\left[ (D_W)_{xy} + g (\sigma_x+ i \gamma_5 \vec{\tau}\cdot\vec{\pi}_x)\delta_{xy} \right]}^{\overline{D}(\phi)}q_y\nonumber\\[1ex]
    &\hspace{2cm} -
    \underbrace{\sum_{x,\mu} \phi_x^i\phi_{x+\hat{\mu} }^i + \sum_x \left[\frac{1}{2} \left(2d+m_0^2 \right) \left(\phi_x^i\phi_x^i\right) + \frac{\lambda}{4!}\left(\phi_x^i\phi_x^i\right)^2 \right]}_{S_\phi}\,,
\end{align}
with $\phi = \left(\sigma, \pi_1,  \pi_2,  \pi_3\right)^T$. The bare lattice Yukawa coupling is related to that in the continuum model with $g=h_\phi/2$ and the bare lattice $\phi^4$ coupling obeys $\lambda=3 \lambda_\phi$. The Wilson-Dirac operator reads
\begin{align}
    (D_W)_{xy} = -\frac{1}{2}\sum_\mu \left[\Gamma_{+\mu} \delta_{x+\hat\mu,y} + \Gamma_{-\mu} \delta_{x-\hat\mu,y} - 2 \delta_{xy}\right] + m_q \delta_{xy} \,,
\end{align}
where $\Gamma_{\pm\mu} = (1 \mp \gamma_\mu)$ and $m_q$ is the (degenerate) bare quark mass.
Note that the $\Gamma_{\pm\mu}$ act in Dirac space, and that $D_W$ is diagonal in flavour space. 

Since the fermion fields enter the action quadratically they can be integrated out. This leaves us with an effective action that is only a function of the meson fields and of the determinant of $D_W$.
Henceforth, we shall use $S$ to denote this effective action 
\begin{align}
    S =  S_\phi-\ln\left(\det\left[\overline{D}(\phi) \right] \right) =   S_\phi-{\textrm{Tr}}\left(\ln\left[\overline{D}(\phi) \right] \right)\,.
    \label{eq: effective action}
\end{align}
The trace in \labelcref{eq: effective action} is performed over spacetime $x$, helicity $h$, spin $s$ and flavour indices $f$. 

As already discussed in the introduction, the two-flavour QM model still retains the chiral properties of QCD at low energies: chiral symmetry breaking in vacuum and its restoration at finite temperature and density. This is monitored by the chiral condensate $\langle \bar q q \rangle$, an order parameter for the chiral phase transition. It vanishes for temperatures $T> T_c$ and is non-vanishing for temperatures $T<T_c$ with the critical temperature $T_c$. Moreover, below $T_c$ the pions are massless as they are the Goldstone bosons of chiral symmetry breaking. 

Typical values for UV-cutoffs in the quark-meson model lie around $\Lambda  = 1$ GeV corresponding to a lattice spacing $a=\frac{\pi}{\Lambda}\approx0.62$ fm.
With feasible temporal extents of our simulation between $N_t=4$ and $N_t = 128$ this LEFT allows thus to investigate relatively low temperatures $T= \frac{1}{aN_t}\approx 80$ MeV, for $N_t=4$, and $T\approx 2.5$ MeV, for $N_t=128$ -- in fact lower than common lattice QCD approaches. Accordingly, LEFTs have a limited temperature range due to the upper limit on the cutoff and the lower limit on temporal lattice sites. Nevertheless, via the method of coloured
noise the accessible range of temperatures as well as its resolution can be enhanced, as discussed in the following.

\section{Stochastic quantisation of the quark-meson model}

Stochastic quantisation~\cite{Parisi:1980ys} provides a way of quantising a given theory.
From a numerical perspective, it has two advantages worth highlighting here.
The actual numerical procedure follows straightforwardly from the analytical equations, i.e., the method is its own algorithm; and it allows, via application of coloured noise, imposing a UV cutoff that is smaller than, but independent of, that introduced by the lattice spacing.
This last point will be elaborated on section~\ref{sec.cn}.

Stochastic quantisation works by augmenting the classical fields with an extra, fictions, time dimension $\tau$ often called ``Langevin time''.
Fields are evolved in $\tau$ following the Langevin equation
\begin{align}\label{eq.Langevin}
\frac{\partial \phi^i_x}{\partial \tau} &= -\frac{\partial S}{\partial \phi^i_x} + \eta^i_x(\tau)\,, 
\end{align}
with $\eta$ a white noise field satisfying
\begin{align}
    \langle \eta^i_x(\tau) \rangle &= 0\,,\\
    \langle \eta^i_x(\tau) \, \eta^j_y(\tau') \rangle &= 2 \delta^{ij} \delta_{xy} \, \delta(\tau - \tau')\,.
\end{align}
In the infinite Langevin time limit the fields are distributed according to $e^{-S}$, the same statistical weight found in path-integral quantisation.
Further details can be found, e.g., in~\cite{damgaard_stochastic_1987}.

\subsection{Coloured noise}\label{sec.cn}

If one removes the noise term from the Langevin equation, eq. (\ref{eq.Langevin}), the resulting equation is known as the \textit{gradient flow}.
The latter has as its long time equilibrium distribution the classical equations of motion, i.e., $\partial S / \partial \phi^i_x = 0$.
Quantum fluctuations are brought into the system via the noise term.

By employing coloured noise, see, e.g.,~\cite{pawlowski_cooling_2017}, it is possible to have a hybrid Langevin time evolution, where fluctuations with momentum larger than a certain energy scale $\Lambda$ follow a gradient flow.
Thus, in the long time limit fluctuations with $p^2 > \Lambda^2$ are suppressed, while those below $\Lambda$ contribute to the quantum expectation values. Employing coloured noise leads ultimately to a decoupling of the underlying lattice spacing $a$ and
the momentum cutoff $\Lambda$.

In practice, this is achieved by considering the noise field $\eta_x^i$ in momentum space, applying some regulator function to it, and Fourier transforming it back to position space.
In this work we follow~\cite{pawlowski_cooling_2017} and impose a sharp cutoff of the form $\theta(\Lambda^2 - p^2)$.
Since our theory is defined on the lattice within a finite volume, the available momentum states are both discrete and finite.
In particular, the largest possible momentum in a $d$-dimensional lattice has magnitude $p_{\mathrm{max}} \equiv \pi \sqrt{d}$ in lattice units.
We therefore impose the cutoff as $\Lambda = s p_{\mathrm{max}}$, with $0 \leq s \leq 1$.
That is, $s$ controls the fraction of momentum states that contribute to the quantum expectation values.
Note that our definition of $s$ differs from that of~\cite{pawlowski_cooling_2017}.

Coloured noise has a few interesting applications.
One of them is finding at which energy scale physics changes, if everything else is kept constant. This is easily understood from a renormalisation group perspective since the removal of momentum shells is strongly connected to a Kadanoff block spin step. The physics on a coarse lattice can be matched with that of a finer one by accompanying the change in lattice size with coloured noise.
The RG-step to keep physics constant is described by the scaling, 
\begin{align}
\label{eq: spacing scaling}
    \Lambda&\rightarrow \Lambda'=s^{-1}\Lambda \,,\nonumber\\[1ex]
    a&\rightarrow a'=sa \,,\nonumber\\[1ex]
    N^d &\rightarrow N'^d = \left( s^{-1} N \right)^d  \,,\nonumber\\[1ex]
    p &\rightarrow p'=s^{-1}p\,.
\end{align}
Note that while the lattice size $N$ and the lattice momenta $p$ are transformed, the physical volume $(aN)^d$ is kept constant.

Via application of coloured noise it is also possible to study the continuum behaviour of a lattice theory, including effective theories. 
This can be achieved by keeping the coloured noise cutoff $\Lambda$ constant, while the lattice cutoff $\Lambda_{\mathrm{UV}} \sim 1/a$ is sent to infinity via adjusting the bare simulation parameters and lattice dimensions. In that way, the arising unphysical UV-modes above the effective cutoff $\Lambda$ are removed. The now possible continuum limit in LEFTs brings the advantage of reducing finite
spacing effects and artefacts coming, e.g., from Wilson fermions as doubler-doubler interactions.
It is also possible to change the temperature with this method, by adjusting the temporal extent of the lattice differently from the spatial ones 
\begin{align}
    N_t \times N_x^{d-1} \rightarrow N_t \times (s^{-1} N_x)^{d-1}.
    \label{eq: scaling N_t fixed}
\end{align}
The temperature will rise by a factor of $s^{-1}$ providing an extended range of overall accessible temperatures and a finer resolution.

\section{Results}
\subsection{Two-dimensional O(\texorpdfstring{$4$}~) scalar field}
First we reproduce a Kadanoff block spin step for an O($4$) scalar theory in $2$-dimensional lattices, i.e., the mesonic sector of a $2$-dimensional quark-meson model.
We start with a simulation done on a initial, ``coarse'', lattice, with lattice cutoff $\Lambda_{\mathrm{UV}}^{\mathrm{coarse}}$ and dimensions $N_t = N_x \equiv N = 8$.
Subsequent simulations are performed at larger lattices of size $N^{\mathrm{fine}} > N$.
In order to keep the \textit{physical} dimensions of the lattice the same as the original simulation, the lattice spacing must obey $a^{\mathrm{fine}} = a N / N^{\mathrm{fine}} < a$.
This change in the lattice spacing implies that bare parameters given in lattice units must also change, such that their physical values remain constant.
The finer spacing $a^{\mathrm{fine}}$, however, implies a larger UV cutoff, which allows for higher energy fluctuations that would change physics. These unwanted momentum shells can be removed by imposing a cutoff $\Lambda^{\mathrm{fine}}_{\mathrm{eff}} = s \Lambda^{\mathrm{fine}}_{\mathrm{UV}} = \Lambda^{\mathrm{coarse}}_{\mathrm{UV}}$ via coloured noise. These relations are summarised in~\Cref{tab:scaling}. It is important to keep in mind that the scaling relations used here are based on tree-level renormalisation group equations.

\begin{table}[h]
\centering
\begin{tabular}{lll}
\toprule
\textrm{Parameter}&
\textrm{Coarse lattice}&
\textrm{Finer lattices}\\
\hline
UV cutoff & $\Lambda_{\mathrm{UV}}^{\mathrm{coarse}}$ &
$\Lambda_{\mathrm{UV}}^{\mathrm{fine}} \equiv s^{-1}\Lambda_{\mathrm{UV}}^{\mathrm{coarse}} \geq \Lambda_{\mathrm{UV}}^{\mathrm{coarse}}$ \\
Effective cutoff &
$\Lambda_{\mathrm{eff}}^{\mathrm{coarse}} = \Lambda_{\mathrm{UV}}^{\mathrm{coarse}}$ &
$\Lambda_{\mathrm{eff}}^{\mathrm{fine}} \equiv
s\Lambda_{\mathrm{UV}}^{\mathrm{fine}}
= \Lambda_{\mathrm{UV}}^{\mathrm{coarse}}$ \\
$($Mass$)^2$ & $a^2 m_0^2$ & $s^2(a^2 m_0^2)$ \\
Coupling & $a^2 \lambda$ & $s^2 (a^2 \lambda)$ \\
Lattice volume & $N_t \times N_x$ & $(s^{-1} N_t) \times (s^{-1} N_x)$\\
\bottomrule
\end{tabular}
\caption{\label{tab:scaling}
Scaling of lattice bare parameters used in order to keep physics constant when going from a lattice with coarse spacing to another with finer spacing in $d=2$.}
\end{table}

\vspace{30mm}
\begin{figure}[htp!]
  \centering
\includegraphics[width=0.8\textwidth]{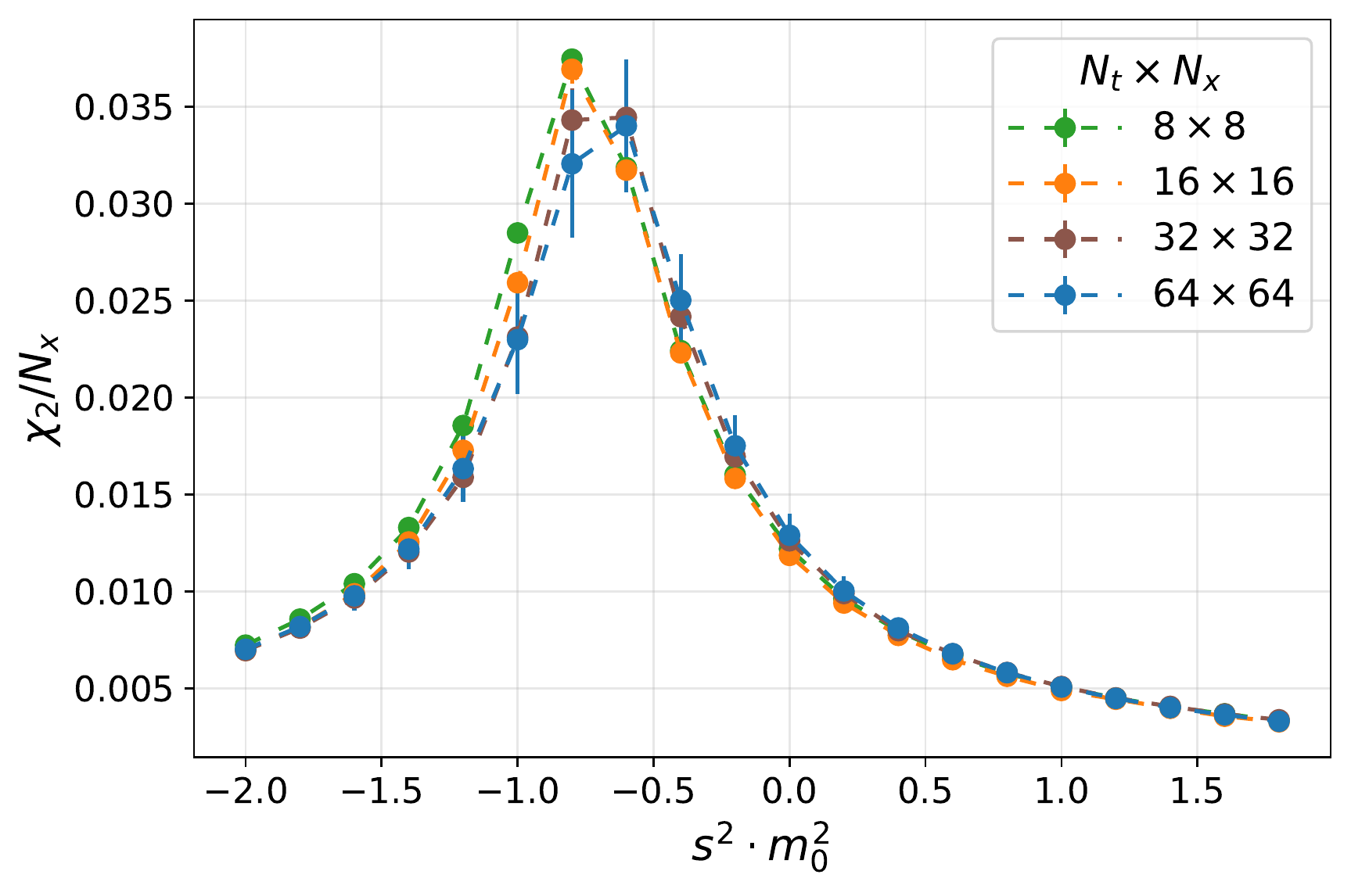}
  \caption{Magnetic susceptibility as function of the rescaled bare mass for lattices of constant physical volume but different numbers of sites.
  Parameters in lattice units: $\lambda_0 = 2.0$.
  \label{fig:constant_physics_b}
  }
\end{figure}

\Cref{fig:constant_physics_b} presents the magnetic susceptibility as a function of the rescaled bare mass in lattice units. The plot shows that the susceptibility remains qualitatively constant as the lattice is made finer, i.e., as $s$ decreases and $N_x, N_t$ increase with the products $s N_t$, $s N_x$ constant. Masses and couplings have been rescaled according to the relations in \Cref{tab:scaling}.
Some discrepancy is noticeable around the peak of the susceptibility which is related to an increase of the autocorrelation time. Note that there is no phase transition here: the peaks mark the crossover between broken and symmetric phases. Similar results have been observed in~\cite{pawlowski_cooling_2017}, where a O($1$) scalar field has been used.

\begin{figure}[htp!]
  \centering
    \includegraphics[width=0.8\textwidth]{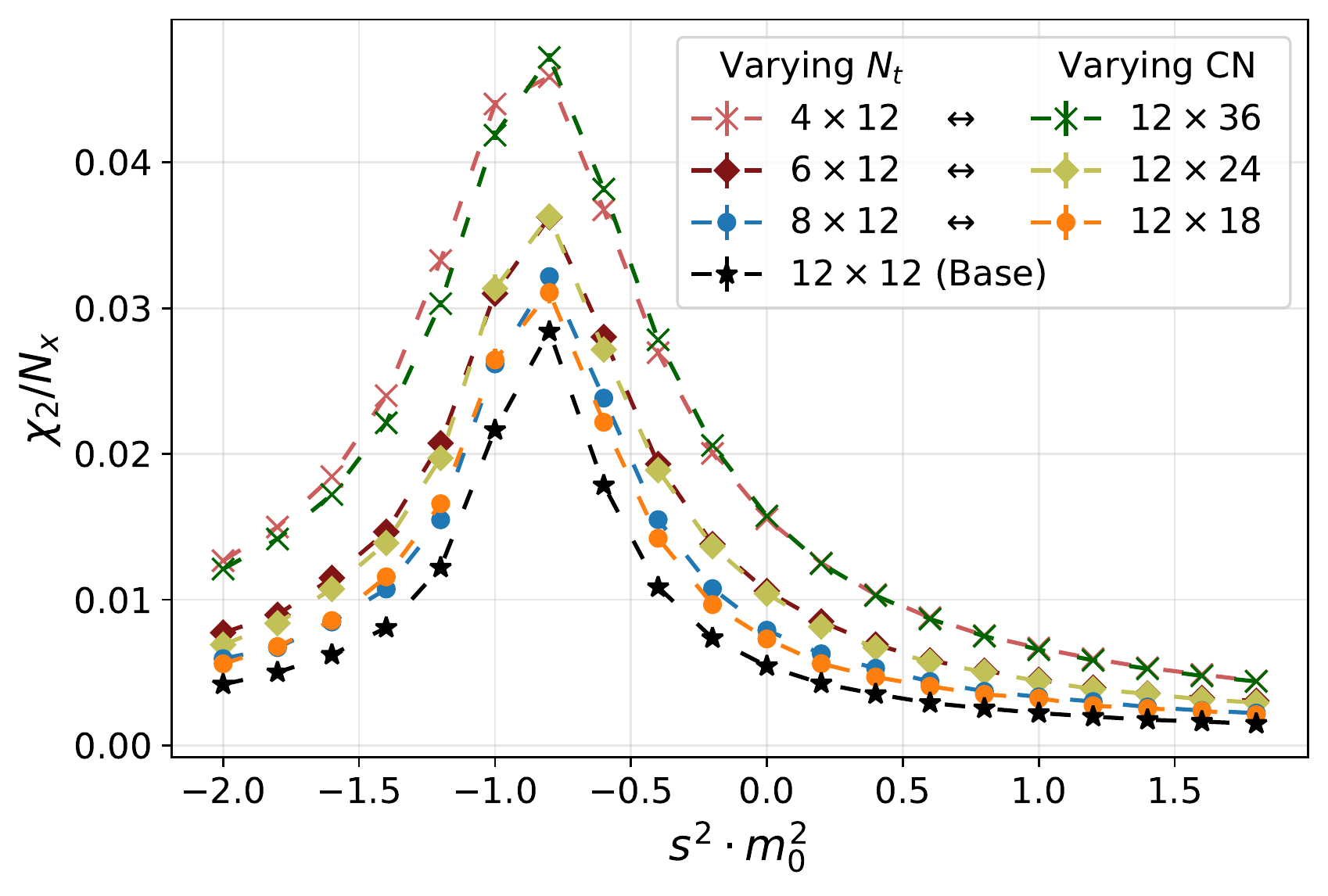}
  \caption{ Magnetic susceptibility as function of the rescaled bare mass for lattices of different temperatures.
  The figure compares changing $T$ via the traditional method of varying $N_t$ and via application of coloured noise.
  Points with the same symbols identify the same temperature.
  Parameters in lattice units: $\lambda_0 = 2.0$.
  \label{fig:change_T}}
\end{figure}

In \Cref{fig:change_T} we demonstrate, that coloured noise can be utilised to perform simulations at different temperatures. Instead of varying the temporal extent the temperature can be controlled by keeping $N_t$ fixed while $N_x$, $a m_0$, $a^2 \lambda$, and $\Lambda_{\mathrm{eff}}$ are rescaled according to~\Cref{tab:scaling}. The figure displays two sets of simulations, one where the temperature has been changed using the traditional method of varying $N_t$ and the other where coloured noise has been employed.
The agreement between the two data sets demonstrates that both methods are equivalent.

\begin{figure}
\centering
\includegraphics[width=0.9\linewidth]{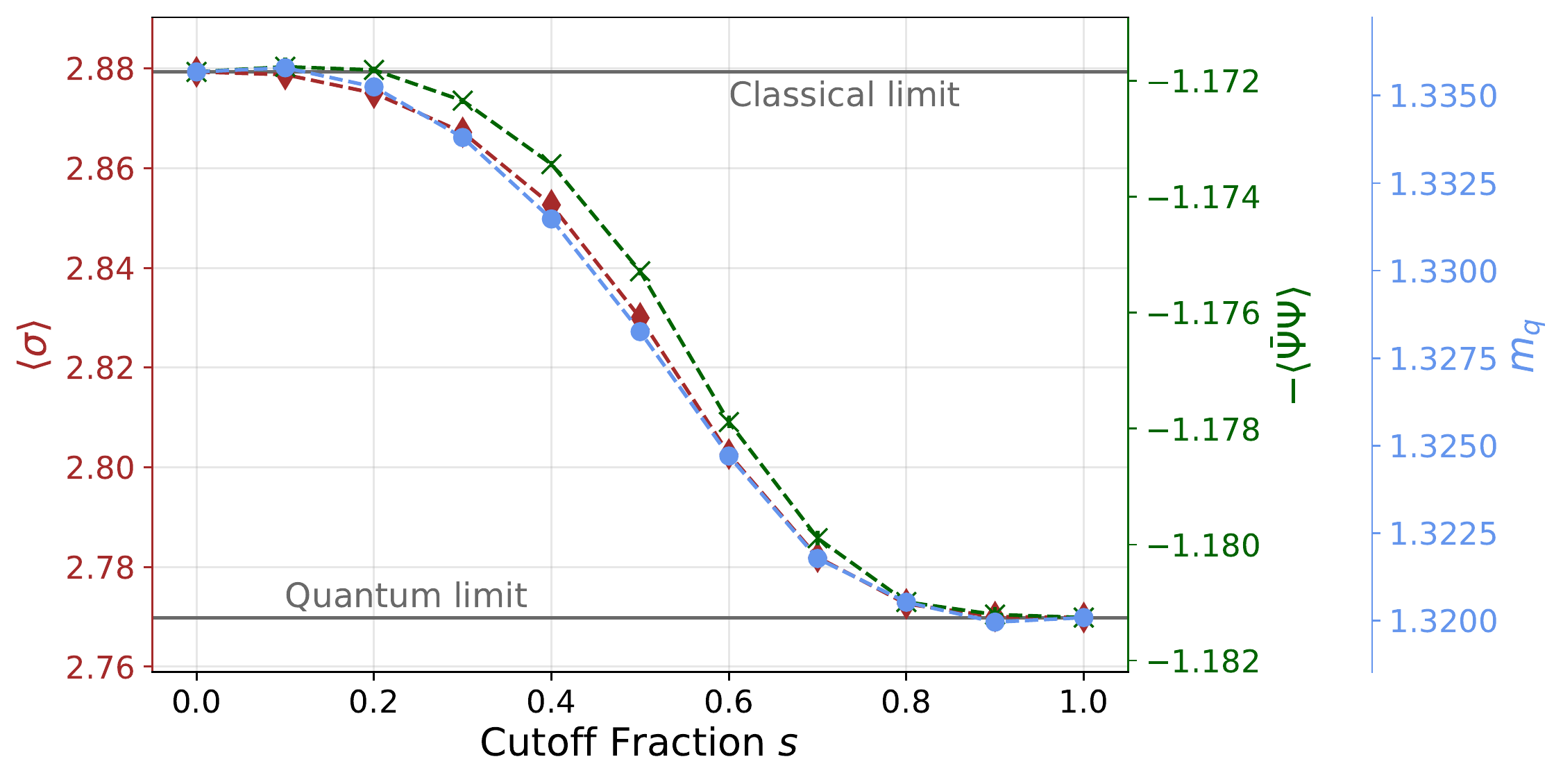}
\caption[Interpolation between classical and quantum theory by employing coloured noise to the quark-meson model]{\label{fig:cutoff_scan_QMM} Interpolation between the classical and quantum limit via varying the cutoff for the quark-meson model. The vacuum expectation value of $\sigma$, the chiral condensate and the physical quark mass are plotted over the 
cutoff fraction $s = \Lambda_{\mathrm{eff}} / \Lambda_{\mathrm{UV}}$.
Note the minus sign in front of the chiral condensate.
Parameters in lattice units: $m_0^2=-1.0, \; \lambda_0 =1.0, \; m_q^0=0.1 \; g=0.5$. Lattice size: $64 \times 8^3$.}
\end{figure}

\subsection{Quark-meson model}
Through the tuning of the effective cutoff $\Lambda_{\mathrm{eff}}$ imposed via coloured noise, certain momentum modes are effectively removed from the simulations.
In the extreme case of $\Lambda_{\mathrm{eff}} \to 0$ one recovers the classical theory. \Cref{fig:cutoff_scan_QMM} shows this interpolation between classical and quantum results for the expectation value of $\sigma$, the chiral condensate, and the quark mass, all measured in lattice units.
Note that these quantities are all related~\cite{Ayala_2021, petropolous}, which is also evident by their qualitatively similar behaviour.
They are plotted as functions of the cutoff fraction $s = \Lambda_{\mathrm{eff}} / \Lambda_{\mathrm{UV}}$.
From right to left, one can see in the plot how the measurements change as the effective cutoff is reduced and higher momentum quantum fluctuations are removed, eventually reaching the classical limit.

\section{Summary and outlook}
We have presented results of lattice simulations of the two-dimensional O($4$) model and the four-dimensional quark-meson model in the presence of an effective momentum cutoff imposed via coloured noise.
Our results demonstrate that coloured noise has various applications, such as investigating the continuum behaviour of effective theories, enabling simulations at different temperatures, and finding at which energy scale the model under consideration changes behaviour.

Many exciting research directions lie ahead.
Simulations of the quark-meson model with coloured noise opens the door for lattice investigations of its phase structure, with higher resolution along the temperature axis and also its continuum behaviour at fixed UV cutoff.
This can then be compared with lattice QCD results at energy scales where the quark-meson model is applicable.
Furthermore, it also provides benchmark results for fRG computations. We hope to report on respective results in the near future.


\acknowledgments
This work is funded by the Deutsche Forschungsgemeinschaft (DFG, German Research Foundation) under Germany's Excellence Strategy EXC 2181/1 - 390900948 (the Heidelberg STRUCTURES Excellence Cluster) and under the Collaborative Research Centre SFB 1225 (ISOQUANT). We also acknowledge support by the state of Baden-W\"urttemberg through bwHPC.



\bibliographystyle{JHEP}
\bibliography{bib}

\end{document}